\documentclass[reprint,noshowpacs,noshowkeys,prd,balancelastpage,nofootinbib]{revtex4}
\usepackage[utf8]{inputenc}
\usepackage{color}
\usepackage{amsmath}
\usepackage{amssymb}
\usepackage[unicode=true,
 bookmarks=false,
 breaklinks=false,pdfborder={0 0 1},backref=section,colorlinks=true]
 {hyperref}
\hypersetup{
 linkcolor=purple,urlcolor=purple,citecolor=blue}

\makeatletter
\@ifundefined{textcolor}{}
{%
 \definecolor{BLACK}{gray}{0}
 \definecolor{WHITE}{gray}{1}
 \definecolor{RED}{rgb}{1,0,0}
 \definecolor{GREEN}{rgb}{0,1,0}
 \definecolor{BLUE}{rgb}{0,0,1}
 \definecolor{CYAN}{cmyk}{1,0,0,0}
 \definecolor{MAGENTA}{cmyk}{0,1,0,0}
 \definecolor{YELLOW}{cmyk}{0,0,1,0}
}

\usepackage{amsfonts}
\usepackage{mdframed}
\usepackage{footnote}
\usepackage{float}
\usepackage[font=footnotesize,it]{caption}
\usepackage{tikz}
\usepackage{tikz-3dplot}

\makeatother

\begin{document}
\title{Higher Dimensional Particle Model in Pure Lovelock Gravity}
\author{S. Danial Forghani}
\email{danial.forghani@emu.edu.tr}

\affiliation{Faculty of Engineering, Final International University, Kyrenia, North
Cyprus via Mersin 10, Turkey}
\author{S. Habib Mazharimousavi}
\email{habib.mazhari@emu.edu.tr}

\affiliation{Department of Physics, Faculty of Arts and Sciences, Eastern Mediterranean
University, Famagusta, North Cyprus via Mersin 10, Turkey}
\author{Mustafa Halilsoy}
\email{mustafa.halilsoy@emu.edu.tr}

\affiliation{Department of Physics, Faculty of Arts and Sciences, Eastern Mediterranean
University, Famagusta, North Cyprus via Mersin 10, Turkey}
\begin{abstract}
In this paper, based on the thin-shell formalism, we introduce a classical
model for particles in the framework of $n+1-$dimensional $\left[\frac{n}{2}\right]$-order
pure Lovelock gravity. In particular, we construct a spherically symmetric
particle of radius $a$ whose inside is a flat Minkowski spacetime
while its outside is charged pLG solution. Knowing that in $n+1-$dimensional
spherically symmetric Einstein gravity ($R$-gravity) such a particle
model cannot be constructed, as we have discussed first, provides
the main motivation for this study. In fact, it is the richness of
Lovelock parameters that provides such a particle construction possible.
On the thin-shell, the energy-momentum components are chosen to vanish,
yet their normal derivatives are non-zero.
\end{abstract}
\date{\today}
\maketitle

\section{Introduction}

Historically, serious attempts for constructing a geometrical framework
within which a particle model can be defined and studied initiated
in 1960s and 1970s \citep{Wheeler1,Vilenkin1}. However, there was
no outstanding achievements mainly because, at the time, the thin-shell
formalism was only developed for Einstein's general relativity whose
solutions were too limited for such a construction. Nowadays, thin-shell
formalism is accessible for many modified theories of gravity such
as $f\left(R\right)$ and Lovelock gravities \citep{Senovilla1,Davis1}.
Accordingly, there has been new attempts to construct such particle
models in modified theories of gravity \citep{Zaslavskii1,Mazharimousavi1,Forghani2}.
The success of these recent studies motivated us to investigate the
likelihood of such models in a rather interesting theory of gravity,
namely the \textit{pure} Lovelock gravity (pLG). This terminology
was used for the first time by Kastor and Mann \citep{Kastor1} and
has been developed by Cai, \textit{et al.} \citep{Cai1,Cai2}, Dadhich,
\textit{et al.} \citep{Dadhich1,Dadhich2,Dadhich3,Dadhich4,Dadhich5,Dadhich6,Dadhich7,Gannouji1,Camanho1,Chakraborty1,Dadhich8}
and others \citep{Mirza1,Concha1,Concha2,Toledoa1,Toledoa2}. The
interesting fact about pLG is that it admits non-degenerate vacua
in even dimensions and unique non-degenerate dS and AdS vacua in odd
dimensions \citep{Cai1}. Moreover, the corresponding black hole solutions
are asymptotically indistinguishable from the ones in Einstein gravity
\citep{Dadhich8}. This similar asymptotic behavior of two theories
seems to extend also to the level of the dynamics and a number of
physical degrees of freedom in the bulk \citep{Dadhich5}. We recall
from the original Lovelock theory that the action of the $n+1-$dimensional
$\left[\frac{n}{2}\right]$-order Lovelock gravity is given by \citep{Lovelock1}
\begin{equation}
I_{\text{LG}}=\int d^{n+1}x\sqrt{-g}\left(\sum_{i=0}^{\left[\frac{n}{2}\right]}\alpha_{i}\mathcal{L}_{i}+\mathcal{L}_{matter}\right)\label{eq:1}
\end{equation}
in which $\left[\frac{n}{2}\right]$ stands for the integral part
of the $\frac{n}{2},$ $\alpha_{i}$ are the $i^{th}-$order Lovelock
parameters, and the Euler densities of a $2i$-dimensional manifold
are given by
\begin{equation}
\mathcal{L}_{i}=\frac{1}{2^{i}}\delta_{\mu_{1}\nu_{1}...\mu_{i}\nu_{i}}^{\alpha_{1}\beta_{1}...\alpha_{i}\beta_{i}}\prod\limits _{s=1}^{i}R_{\text{ \ \ \ \ \ \ \ }\alpha_{i}\beta_{i}}^{\mu_{i}\nu_{i}},\label{eq:2}
\end{equation}
where the generalized Kronecker delta $\delta$ is defined as the
antisymmetric product
\begin{equation}
\delta_{\mu_{1}\nu_{1}...\mu_{i}\nu_{i}}^{\alpha_{1}\beta_{1}...\alpha_{i}\beta_{i}}=i!\delta_{\lbrack\mu_{1}}^{\alpha_{1}}\delta_{\nu_{1}}^{\beta_{1}}...\delta_{\mu_{n}}^{\alpha_{n}}\delta_{\nu_{n}]}^{\beta_{n}}.\label{eq:3}
\end{equation}
On the other hand, in pLG the action is expressed as
\begin{equation}
I_{\text{pLG}}=\int d^{n+1}x\sqrt{-g}\left(\alpha_{0}+\alpha_{p}\mathcal{L}_{p}+\mathcal{L}_{\text{matter}}\right)\label{eq:4}
\end{equation}
in which $\alpha_{0}$ represents the cosmological constant and $1\leq p\leq\left[\frac{n}{2}\right]$.
For instance, the Einstein $R$-gravity has $p=1,$ $\alpha_{1}=1$
and $\mathcal{L}_{1}=R$ which is the particular case of the pLG applicable
in all dimensions. For $p=2$, one finds the pure Gauss-Bonnet (pGB)
gravity applicable in $n+1\geq5.$ Finally, in this paper, $p=3$
represents pure third order Lovelock gravity (pTOLG) which is valid
for $n+1\geq7.$ Herein, we give a general formalism in the three
different pure Lovelock theories mentioned above, i.e., the pure Einstein's
gravity, pGB and pTOLG. It will be shown that in specific cases where
the inner and outer spacetimes of the chosen thin-shell boundary admit
identical Lovelock parameters i.e., $\alpha_{p}^{+}=\alpha_{p}^{-}$,
the junction conditions result the same in terms of the metric functions
and their first derivatives, irrespective of the order of the Lovelock
term. These junction conditions are simply the continuity of the bulk's
metric function and its first derivatives across the thin-shell which
is the surface of the particle. These are the continuity of the first
and second fundamental forms of the surface.

In the next section (Sec. \ref{SecII}), we separately study our particle
model for three different orders of pLG. These studies are followed
by our conclusion brought in the last section (Sec. \ref{Sec III}).
All over the manuscript, we use the unit convention $4\pi\epsilon_{0(n+1)}=8\pi G_{(n+1)}=\hbar=c=1$.

\section{\label{SecII}Particle model in pure Lovelock gravity}

\subsection{Pure Einstein gravity}

Our spherically symmetric line elements for the exterior and interior
of the particle are chosen to be
\begin{equation}
ds^{2}=-f_{\pm}\left(r\right)dt^{2}+f_{\pm}^{-1}\left(r\right)dr^{2}+r^{2}\left(d\theta^{2}+\sin^{2}\theta d\phi^{2}\right),\label{eq:1-1}
\end{equation}
and in this section we construct an $n+1$-dimensional chargeless
particle model in pure Einstein $R$-gravity. We assume a static timelike
spherical shell of radius $r=a$ such that its exterior and interior
metric functions are $f_{+}\left(r_{+}\right)$ and $f_{-}\left(r_{-}\right),$
respectively (See Appendix A for a summary on the thin-shell formalism).
The standard Israel junction conditions \citep{Darmois1,Israel1}
imply that the energy density $\sigma$ and the tangential pressure
$p$ of the shell are given by \citep{Mehdizadeh1} 
\begin{equation}
\sigma=-\frac{n-1}{a}\left(\sqrt{f_{+}}-\sqrt{f_{-}}\right)\label{eq:5}
\end{equation}
and
\begin{equation}
p=\frac{1}{2}\left(\frac{f_{+}^{\prime}}{\sqrt{f_{+}}}-\frac{f_{-}^{\prime}}{\sqrt{f_{-}}}\right)-\frac{n-2}{n-1}\sigma,\label{eq:6}
\end{equation}
respectively. Herein, $a$ is the radius of the surface of the particle
and a prime ($^{\prime}$) stands for a total derivative with respect
to the radial coordinates. Also, all the metric functions and their
respective derivatives are evaluated at $a$. For a particle, one
expects $\sigma=0$ and $p=0$ simultaneously \citep{Vilenkin1,Zaslavskii1}
which result in the bulk's metric function and its first derivative
to be continuous across the surface of the particle, i.e. $\left(f_{+}=f_{-}\right)_{\Sigma}$
and $\left(f_{+}^{\prime}=f_{-}^{\prime}\right)_{\Sigma}$. In other
words, continuity of the first and second fundamental forms across
the thin-shell correspond to having the metric function and its first
derivative continuous on the surface of the particle. Let us comment
that $\sigma=0=p$ is not trivially satisfied on the shell. We may
yet have the higher and normal derivatives of $\sigma$ and $p$ non-zero
which amount to the existence of shell matter. Further, any perturbation
of the shell will distort the equilibrium condition of stress-energy,
so that we shall have $\sigma\neq0\neq p$ after the perturbation.
Hence, we are very much restricted in choosing the spacetimes inside
and outside of the particle. For instance, a Minkowski inner flat
and an outer Schwarzschild spacetimes do not satisfy the two conditions.
Also, an inner Minkowski flat and an outer Reissner-Nordström (RN)
can not be matched, whereas a cloud of strings of the form \citep{Letelier1}
\begin{equation}
f_{-}=1-\frac{2\mu}{\left(n-1\right)r^{n-3}}\label{eq:7}
\end{equation}
can be matched to the RN spacetime with the metric function 
\begin{equation}
f_{+}=1-\frac{2m}{r^{n-2}}+\frac{q^{2}}{r^{2\left(n-2\right)}}.\label{eq:8}
\end{equation}
Herein, $\mu$ is an integration constant associated with the cloud
of strings, $m$ is a mass parameter (proportional to the physical
mass) and $q$ is the charge parameter (proportional to the physical
total charge). Applying $\left(f_{+}=f_{-}\right)_{\Sigma}$ and $\left(f_{+}^{\prime}=f_{-}^{\prime}\right)_{\Sigma}$
yields 
\begin{equation}
\frac{m}{q^{2}}=\frac{n-1}{2a^{n-2}},\label{eq:9}
\end{equation}
and
\begin{equation}
\frac{\mu}{q^{2}}=\frac{\left(n-1\right)\left(n-2\right)}{2a^{n-1}}.\label{eq:10}
\end{equation}
Specifically, in $4$-dimensional spacetime ($n=3$) one finds $a=\frac{q^{2}}{m}$
and $\mu=\left(\frac{m}{q}\right)^{2}$. Furthermore, in $4$-dimensional
spacetime the Newton's potential inside the shell becomes a constant,
i.e. $f_{-}=1+2\Phi_{-}=1-\mu$, which leads to $\Phi_{-}=-\frac{\mu}{2}$.
In addition, the exterior potential of the particle is given by $f_{+}=1+2\Phi_{+}=1-\frac{2m}{r}+\frac{q^{2}}{r^{2}}$,
or explicitly $\Phi_{+}=-\frac{m}{r}+\frac{q^{2}}{2r^{2}}$. On the
surface, we trivially have $\Phi_{-}=$ $\Phi_{+}=-\frac{\mu}{2}$,
which is in agreement with our understanding of Newtonian potential
inside and outside a spherical shell in classical mechanics. We should
admit, however, that although the constant Newtonian potential seems
a better choice than a flat Minkowski for the particle's interior,
such a spacetime is singular due to the singular nature of the energy-momentum
tensor of the string cloud
\begin{equation}
T_{\mu}^{\nu}=diag\left(\frac{\mu}{r^{2}},\frac{\mu}{r^{2}},0,0\right).\label{eq:11}
\end{equation}
In higher-dimensional particle models without singularity at the center,
one may think of a(n) (Anti-)de Sitter spacetime with the cosmological
constant $\Lambda=\pm\frac{n\left(n-1\right)}{2\ell^{2}}$ as the
interior of the particle in the form
\begin{equation}
f_{-}=1-\frac{r^{2}}{\ell^{2}},\label{eq:12}
\end{equation}
and an RN spacetime as the exterior of the particle, given in Eq.
(\ref{eq:8}). Upon matching the two metrics, we obtain 
\begin{equation}
\frac{m}{q^{2}}=\frac{n-1}{na^{n-2}},\label{eq:13}
\end{equation}
and
\begin{equation}
q^{2}\ell^{2}=\frac{n}{\left(n-2\right)}a^{2\left(n-1\right)}.\label{eq:14}
\end{equation}
Again in the specific $4$-dimensional case ($n=3$), we find $a=\frac{2}{3}\left(\frac{q^{2}}{m}\right)$
and $\ell^{2}=\frac{16}{27}\frac{q^{6}}{m^{4}}$. By reversing these
expressions, we obtain the mass and the charge of a particle in terms
of the geometric parameters of the theory. We must add that, in all
cases the particle's radius $a$ should be grater than the radius
of the event horizon of the exterior metric and smaller than the possible
cosmological horizon of the interior metric. These are necessary to
avoid any horizon or singularity inside or outside of the particle.
For instance, again in four dimensions, in order to avoid any horizon
within outer spacetime one has to assume $a>\frac{2}{3}q$.

As the final example for this section, let us consider a global monopole
for the interior spacetime with the metric function 
\begin{equation}
f_{-}=1-2\eta\label{eq:15}
\end{equation}
in an arbitrary dimension, and $f_{+}$ given by Eq. (\ref{eq:8})
for the exterior spacetime. The junction conditions impose $\left(f_{+}=f_{-}\right)_{\Sigma}$
and $\left(f_{+}^{\prime}=f_{-}^{\prime}\right)_{\Sigma}$ which in
turn reveal the radius as
\begin{equation}
a=\left(\frac{q^{2}}{m}\right)^{\frac{1}{n-2}}.\label{eq:16}
\end{equation}
 Note that, the energy-momentum tensor of the global monopole spacetime
which is given by
\begin{equation}
T_{\nu}^{\mu}=-\frac{\eta}{r^{2}}\left(n-2\right)diag\left(n-1,n-1,\underset{n-1\text{ times}}{\underbrace{n-3,...,n-3}}\right),\label{eq:17}
\end{equation}
implies that the spacetime is singular at $r=0$. In $3+1$-dimensions,
however, the spacetime reduces to the cloud of strings of the previous
example with $\rho=-p_{r}=\eta/r^{2}$, and $p_{\theta}=p_{\phi}=0$.
In higher dimensions, if one desires to keep the potential within
the interior spacetime to be constant, the only candidate spacetime
is the global monopole. Unfortunately, such a spacetime admits undesirable
singularity at $r=0$.

\subsection{Pure Gauss-Bonnet gravity}

In \citep{Cai1}, the pLG with cosmological constant has been investigated.
In pGB gravity, the action amounts to
\begin{equation}
I_{\text{pGB}}=\int dx^{n+1}\sqrt{-g}\left(\alpha_{0}+\alpha_{2}\mathcal{L}_{\text{GB}}\right),\text{ }n\geq4\label{eq:18}
\end{equation}
in which $\alpha_{0}$ is the cosmological constant, $\alpha_{2}$
is the GB free parameter and 
\begin{equation}
\mathcal{L}_{\text{GB}}=R^{\kappa\lambda\mu\nu}R_{\kappa\lambda\mu\nu}-4R^{\mu\nu}R_{\mu\nu}+R^{2}\label{eq:19}
\end{equation}
 is the GB Lagrangian. In \citep{Cai1}, a spherically symmetric solution
with the line element 
\begin{equation}
ds^{2}=-f\left(r\right)dt^{2}+\frac{1}{f\left(r\right)}dr^{2}+r^{2}d\Omega_{n-1}^{2}\label{eq:20}
\end{equation}
 is considered such that the metric function is 
\begin{equation}
f\left(r\right)=1\pm\frac{r^{2}}{\sqrt{\tilde{\alpha}_{2}}}\sqrt{\frac{2M}{\left(n-1\right)\Sigma_{n-1}r^{n}}+\frac{1}{\ell^{2}}.}\label{eq:21}
\end{equation}
Here, $M$ is the mass of the solution, $\Sigma_{n-1}=\frac{2\pi^{n/2}}{\Gamma\left(\frac{n}{2}\right)}$
is the volume of $\left(n-1\right)$-sphere \citep{Toledoa2}, $\tilde{\alpha}_{2}$
is given by
\begin{equation}
\tilde{\alpha}_{2}=\left(n-2\right)\left(n-3\right)\alpha_{2},\label{eq:22}
\end{equation}
and 
\begin{equation}
\frac{1}{\ell^{2}}=-\frac{\alpha_{0}}{n\left(n-1\right)}\label{eq:23}
\end{equation}
is related to the cosmological constant $\alpha_{0}$. In particular,
for $n+1=7$ one finds
\begin{equation}
f\left(r\right)=1\pm\frac{r^{2}}{\sqrt{\tilde{\alpha}_{2}}}\sqrt{\frac{2M}{5\pi^{3}r^{6}}+\frac{1}{\ell^{2}}},\label{eq:24}
\end{equation}
in which the parameters, upon choosing the ($+$) sign, are connected
with $\frac{1}{\ell^{2}}=-\frac{\alpha_{0}}{30}$ and $\tilde{\alpha}_{2}=12\alpha_{2}.$
Note that an interesting property of the $7$-dimensional pGB theory
without cosmological constant is that its potential, from Eq. (\ref{eq:24}),
gives the same fall-off as in the $4$$-$dimensional Einstein gravity
\citep{Dadhich2} i.e., 
\begin{equation}
\Phi\simeq\pm\frac{m}{r}\label{eq:25}
\end{equation}
where $m=\frac{1}{2\sqrt{\tilde{\alpha}_{2}}}\sqrt{\frac{2M}{5\pi^{3}}}$.

Next, let us consider a particle of radius $a$ whose inner and outer
spacetimes are solutions in pGB. The generalized Israel junction condition
must be applied at the surface of the particle where the two incomplete
spacetimes are glued. Based on the generalized Israel junction conditions
\citep{Davis1}, without assuming $\frac{dt_{+}}{d\tau}=\frac{dt_{-}}{d\tau}$,
one simply finds the induced metric on the shell to be given by 
\begin{equation}
ds^{2}=-d\tau^{2}+a^{2}d\Omega_{n-1}^{2},\label{eq:26}
\end{equation}
where $a=a\left(\tau\right)$ is a function of the proper time on
the shell. The surface energy-momentum tensor $S_{b}^{a}=\left(-\sigma,\underset{\text{\ensuremath{n-1}-times}}{\underbrace{p,...,p}}\right)$
is expressed as \citep{Forghani1}
\begin{equation}
-S_{b}^{a}=2\left[\alpha_{2}\left(3J_{b}^{a}-J\delta_{b}^{a}\right)\right]_{-}^{+}\label{eq:27}
\end{equation}
in which $J_{b}^{a}$ and $J$ are defined as
\begin{equation}
J_{b}^{a}=diag\left(-\frac{2!}{3}\left\{ \sum_{s=0}^{2}\frac{\left(-1\right)^{s}}{n}\binom{n}{s}\left[sK_{\tau}^{\tau}+\left(n-s\right)K_{\theta}^{\theta}\right]\left(K_{\theta}^{\theta}\right)^{s-1}\left(K_{b}^{a}\right)^{5-s}\right\} \right)\label{eq:28}
\end{equation}
and
\begin{equation}
J=J_{a}^{a}\label{eq:29}
\end{equation}
respectively. Here, $K_{\tau}^{\tau}$ and $K_{\theta}^{\theta}=K_{\theta_{1}}^{\theta_{1}}=K_{\theta_{2}}^{\theta_{2}}=...=K_{\theta_{n-1}}^{\theta_{n-1}}$\ are
the components of the extrinsic curvature tensor associated with the
time and angular coordinates of the timelike hyperplane (surface of
the particle). An explicit calculation reveals 
\begin{equation}
\sigma=\frac{2\left(n-1\right)}{3a^{3}}\left[\tilde{\alpha}_{2}\sqrt{f}\left(f-3\right)\right]_{-}^{+}\label{eq:30}
\end{equation}
and
\begin{equation}
p=-\frac{1}{a^{2}}\left[\frac{\tilde{\alpha}_{2}\left(f-1\right)f^{\prime}}{\sqrt{f}}\right]_{-}^{+}+\frac{n-4}{n-1}\sigma.\label{eq:31}
\end{equation}
Having considered specific metric functions $f_{\pm}$, and also $\tilde{\alpha}_{2}^{-}$
and $\tilde{\alpha}_{2}^{+}$ for the interior and exterior spacetimes,
respectively, the conditions $\sigma=0$ and $p=0$ yield
\begin{equation}
\tilde{\alpha}_{2}^{+}\sqrt{f_{+}}\left(f_{+}-3\right)-\tilde{\alpha}_{2}^{-}\sqrt{f_{-}}\left(f_{-}-3\right)=0\label{eq:32}
\end{equation}
and
\begin{equation}
\tilde{\alpha}_{2}^{+}\sqrt{f_{-}}f_{+}^{\prime}\left(f_{+}-1\right)-\tilde{\alpha}_{2}^{-}\sqrt{f_{+}}f_{-}^{\prime}\left(f_{-}-1\right)=0.\label{eq:33}
\end{equation}
For $\tilde{\alpha}_{2}^{+}=\tilde{\alpha}_{2}^{-}$, one may consider
the trivial solution of Eq. (\ref{eq:32}) as $\left(f_{+}=f_{-}\right)_{\Sigma}$,
and consequently, Eq. (\ref{eq:33}) yields $\left(f_{+}^{\prime}=f_{-}^{\prime}\right)_{\Sigma}$.
For $\tilde{\alpha}_{2}^{+}\neq\tilde{\alpha}_{2}^{-}$ Eq. (\ref{eq:32})
implies
\begin{equation}
\tilde{\alpha}_{2}^{-}=\frac{\sqrt{f_{+}}\left(f_{+}-3\right)}{\sqrt{f_{-}}\left(f_{-}-3\right)}\tilde{\alpha}_{2}^{+}\label{eq:34}
\end{equation}
which upon inserting into Eq. (\ref{eq:33}) we obtain
\begin{equation}
\left(f_{-}-3\right)\left(f_{+}-1\right)f_{-}f_{+}^{\prime}+\left(f_{+}-3\right)\left(f_{-}-1\right)f_{+}f_{-}^{\prime}=0.\label{eq:35}
\end{equation}
This is the only constraint on the metric functions and their first
derivatives to be satisfied on $\Sigma.$ As an example, let us consider
$f_{-}=1$, and the positive branch of the solution in Eq. (\ref{eq:21})
as $f_{+\text{ }}$.Note that upon choosing $f_{-}=1$, we implicitely
assumed $\alpha_{0}^{-}=0$, whereas the outer cosmological constant
is generally non-zero. This jump in cosmological constant could be
due to the fine structure of the particle and is mathematically allowed,
as far as the junction conditions are concerned. Multiple spherical
branes with a position-dependent cosmological constant are studied
in \citep{Jardim1}. Considering $\tilde{\alpha}_{2}^{+}\neq\tilde{\alpha}_{2}^{-}$,
from Eq. (\ref{eq:35}) and Eq. (\ref{eq:34}) we obtain 
\begin{equation}
M=\frac{2\left(n-1\right)\Sigma_{n-1}a^{n}}{\ell^{2}\left(n-4\right)}\label{eq:36}
\end{equation}
and 
\begin{equation}
\tilde{\alpha}_{2}^{-}=\frac{\tilde{\alpha}_{2}^{+}}{2}\left(2-\lambda_{2}\right)\sqrt{1+\lambda_{2}}\label{eq:37}
\end{equation}
in which
\begin{equation}
\lambda_{2}=\frac{a^{2}}{\sqrt{\tilde{\alpha}_{2}^{+}}}\sqrt{\frac{n}{\left(n-4\right)\ell^{2}}}.\label{eq:38}
\end{equation}
To complete the argument, let us add that, in the case where $\tilde{\alpha}_{2}^{-}=\tilde{\alpha}_{2}^{+}$
the junction conditions simply reduce to the one in $R$$-$gravity,
i.e., the continuity of the bulk's metric function and its first derivative.

\subsection{Pure third-order Lovelock gravity}

Similar to the previous section, in this section we shall give a particle
model in pure TOLG. Following \citep{Cai1,Cai2}, the corresponding
action is given by 
\begin{equation}
I_{\text{pTOLG}}=\int dx^{n+1}\sqrt{-g}\left(\alpha_{0}+\alpha_{3}\mathcal{L}_{\text{TOLG}}\right),\text{ }n\geq6\label{eq:39}
\end{equation}
in which $\alpha_{0}$ stands for the cosmological constant, $\alpha_{3}$
is the third order Lovelock free parameter and 
\begin{multline}
\mathcal{L}_{\text{TOLG}}=2R^{\kappa\lambda\rho\sigma}R_{\rho\sigma\mu\nu}R_{\hspace{0.3cm}\kappa\lambda}^{\mu\nu}+8R_{\hspace{0.3cm}\kappa\lambda}^{\mu\nu}R_{\hspace{0.3cm}\nu\rho}^{\kappa\sigma}R_{\hspace{0.3cm}\mu\sigma}^{\lambda\rho}+24R^{\kappa\lambda\mu\nu}R_{\mu\nu\lambda\rho}R_{\hspace{0.15cm}\kappa}^{\rho}\\
+3RR^{\kappa\lambda\mu\nu}R_{\kappa\lambda\mu\nu}+24R^{\kappa\lambda\mu\nu}R_{\mu\kappa}R_{\nu\lambda}+16R^{\mu\nu}R_{\nu\sigma}R_{\hspace{0.15cm}\mu}^{\sigma}-12RR^{\mu\nu}R_{\mu\nu}+R^{3}\label{eq:40}
\end{multline}
is the third order Lovelock Lagrangian. In \citep{Cai1}, after supplementing
the Lagrangian with a Maxwell term, an $n+1$$-$dimensional static
spherically symmetric solution with electric charge is found whose
line element is given by Eq. (\ref{eq:21}), with the metric function
written as
\begin{equation}
f\left(r\right)=1+\frac{r^{2}}{\left\vert \tilde{\alpha}_{3}\right\vert ^{1/3}}\sqrt[3]{\frac{2M}{\left(n-1\right)\Sigma_{n-1}r^{n}}-\frac{q^{2}}{r^{2\left(n-1\right)}}+\frac{1}{\ell^{2}}}.\label{eq:41}
\end{equation}
Herein, the integration constants $M$ and $q$ are the usual mass
and electric charge parameters, $\ell^{2}>0$ represents the cosmological
length and $\tilde{\alpha}_{3}=-\left\vert \tilde{\alpha}_{3}\right\vert <0.$
This solution contains two theory parameters i.e., $\tilde{\alpha}_{3}$
and $\ell^{2}$, and two solution parameters which are the integration
constants i.e., $M$ and $q.$

Next, we consider a spherically symmetric particle that consists of
inner and outer spacetimes as solutions of pure TOLG, such as the
one given in (\ref{eq:41}). The associated junction conditions extracted
from \citep{Dehghani1,Dehghani2,Dehghani3} are expressed as \citep{Forghani1}
\begin{equation}
-S_{b}^{a}=\left[3\alpha_{3}\left(5P_{b}^{a}-P\delta_{b}^{a}+\mathcal{L}_{GB}\left(K_{b}^{a}+K\delta_{b}^{a}\right)\right)\right]_{-}^{+},\label{eq:42}
\end{equation}
in which $S_{b}^{a},$and $K_{b}^{a}$ are as before while $K=K_{a}^{a}$
with
\begin{equation}
P_{b}^{a}=diag\left(\frac{4!}{5}\left\{ \sum_{s=0}^{4}\frac{\left(-1\right)^{s}}{n}\binom{n}{s}\left[sK_{\tau}^{\tau}+\left(n-s\right)K_{\theta}^{\theta}\right]\left(K_{\theta}^{\theta}\right)^{s-1}\left(K_{b}^{a}\right)^{5-s}\right\} \right)\label{eq:43}
\end{equation}
and $P=P_{a}^{a}.$ Within explicit calculations we obtain 
\begin{equation}
\sigma=-\frac{\left(n-1\right)}{5a^{5}}\left[\tilde{\alpha}_{3}\sqrt{f}\left(3f^{2}-10f+15\right)\right]_{-}^{+}\label{eq:44}
\end{equation}
and
\begin{equation}
p=\frac{3}{2a^{4}}\left[\frac{\tilde{\alpha}_{3}\left(1-f\right)^{2}f^{\prime}}{\sqrt{f}}\right]_{-}^{+}-\frac{n-6}{n-1}\sigma.\label{eq:45}
\end{equation}
As of the boundary conditions on $\Sigma$, the surface of the particle,
the first fundamental form $h_{ab}$ and the second fundamental form
$K_{ab}$ (see Appendix A) should be continuous. Technically these
are equivalent to $\sigma=0$ and $p=0$ simultaneously. Similar to
the pure GB case, there are two distinct possibilities: i) $\tilde{\alpha}_{3}^{+}=\tilde{\alpha}_{3}^{-}$
which might trivially imply $\left(f_{+}=f_{-}\right)_{\Sigma}$ and
so $\left(f_{+}^{\prime}=f_{-}^{\prime}\right)_{\Sigma}$, and ii)
$\tilde{\alpha}_{3}^{+}\neq\tilde{\alpha}_{3}^{-}$ which results
in the following relation between $\tilde{\alpha}_{3}^{+}$ and $\tilde{\alpha}_{3}^{-}$
\begin{equation}
\tilde{\alpha}_{3}^{-}=\frac{\sqrt{f_{+}}\left(3f_{+}^{2}-10f_{+}+15\right)}{\sqrt{f_{-}}\left(3f_{-}^{2}-10f_{-}+15\right)}\tilde{\alpha}_{3}^{+}\label{eq:46}
\end{equation}
and a constraint condition on the metric functions 
\begin{equation}
f_{+}f_{-}^{\prime}\left(1-f_{-}\right)^{2}\left(3f_{+}^{2}-10f_{+}+15\right)-f_{-}f_{+}^{\prime}\left(1-f_{+}\right)^{2}\left(3f_{-}^{2}-10f_{-}+15\right)=0.\label{eq:47}
\end{equation}
Therefore, for $\tilde{\alpha}_{3}^{+}\neq\tilde{\alpha}_{3}^{-}$,
every considered solution for the inner and outer spacetimes must
satisfy this condition on $\Sigma$. Moreover with $\tilde{\alpha}_{3}^{+}=\tilde{\alpha}_{3}^{-}$
the junction conditions reduce to the one in $R$$-$gravity and also
in pure GB gravity with $\tilde{\alpha}_{2}^{+}=\tilde{\alpha}_{2}^{-}.$

As an example, since we are interested to construct a singularity
free particle model, let us set $f_{-}=1$ (and consequently $f_{-}^{\prime}=0$).
The first condition, i.e. $\sigma=0$, implies
\begin{equation}
\tilde{\alpha}_{3}^{-}=\frac{\tilde{\alpha}_{3}^{+}}{8}\sqrt{f_{+}}\left(3f_{+}^{2}-10f_{+}+15\right).\label{eq:48}
\end{equation}
Considering the second condition, i.e. $p=0$, we find
\begin{equation}
f_{+}^{\prime}\left(1-f_{+}\right)^{2}=0\label{eq:49}
\end{equation}
which admits two possibilities, either $f_{+}=1$ on the surface which
consequently implies $\tilde{\alpha}_{3}^{+}=\tilde{\alpha}_{3}^{-}$
or $f_{+}^{\prime}=0$. Upon considering $f_{+}$ to be the general
charged solution given in (\ref{eq:41}) without the cosmological
constant, the latter equation, i.e., $f_{+}^{\prime}=0$ admits the
relation 
\begin{equation}
\frac{M}{q^{2}}=\frac{\left(n-1\right)\left(n-4\right)\Sigma_{n-1}}{\left(n-6\right)a^{n-2}}\label{eq:50}
\end{equation}
between the mass and the charge of the particle in terms of its radius
and 
\begin{equation}
\tilde{\alpha}_{3}^{-}=\frac{\tilde{\alpha}_{3}^{+}}{8}\sqrt{1+\lambda_{3}}\left(3\lambda_{3}^{2}-4\lambda_{3}+8\right)\label{eq:51}
\end{equation}
in which 
\begin{equation}
\lambda_{3}=a^{2}\left(\frac{\left(n-2\right)q^{2}}{\left\vert \tilde{\alpha}_{3}^{+}\right\vert \left(n-6\right)a^{2\left(n-1\right)}}\right)^{1/3}.\label{eq:52}
\end{equation}
In $7-$dimensional spacetime where $n=6$, however, such a particle
model fails to work and one needs to consider additional theory parameters
such as a cosmological constant. In short, having the third-order
parameters involved in this specific case, provides us a particle
model with mass and charge which could not be made in $R$-gravity.
This shows the rich structure of the LG in any form in constructing
particle models. Once more, we add that to avoid nonphysical particle
models, the spacetime of the particle including inside, on, and outside
the shell have to be singularity- and horizon-free. These are the
additional conditions to be imposed on the radius of any physical
particle.

\section{\label{Sec III}Conclusion}

We employed the cut and paste method to construct a classical particle
model in $n+1-$dimensional spherically symmetric pLG. In general,
for an $n+1-$dimensional particle model, when the Lovelock parameters
of the outer and inner spacetimes are identical, irrespective of the
number of dimensions and the order of pLG, the junction conditions
reduce to the continuity of the metric function and its first derivative
on the surface of the particle. As a result, problem becomes trivial,
meaning that there is no particle model to worry about. On the other
hand, with different Lovelock parameters the scenario changes dramatically
such that depending on what order of Lovelock gravity is considered
one obtains two distinct conditions relating the metric functions
as well as the Lovelock parameters. In pGB and pTOLG gravities, we
have explicitly found these relations. Although in $n+1-$dimensional
$R-$gravity a particle model with flat inside and Schwarzschild or
RN outside are not possible (for $n=3$ see \citep{Zaslavskii1}),
in both pGB and pTOLG such a massive and charged model of particle
become feasible. It is important to perceive that the thin-shell formalism
allows us to choose different Lovelock parameters (or even cosmological
constants) for the inside and outside of the particle as long as we
are using the same gravity theory for the two sides. Eventually, the
mass, charge and radius of the constructed particle models are defined
entirely from the geometrical parameters, which is required from a
geometrical theory.

At this point, It is also worth mentioning that within any theory
that is studied here or elsewhere, there has to exist a factor to
distinguish the interior of the particle from its exterior. In this
study, in particular, this factor has been the Lovelock parameters
(or the cosmological constant in some cases). However, to consider
different particles one must hold the exterior Lovelock parameter
fixed, since for every particle the exterior is the same. Therefore,
it is indeed different interior parameters that may give rise to different
particles. Nonetheless, we also admit that, discontinuous Lovelock
parameters is quite an exotic proposal since they are coupling constants
which by default are constant. This is in analogy with different gravitational
constants $G$ across a hypersurface, which actually has been studied
in \citep{Takamizu1}.

In this study, we considered only a very limited number of of solutions
as for the interior and exterior spacetimes. For example, in most
cases we consider a flat Minkowski spacetime for the interior of the
particle, whereas choosing rather complicated solutions with a non-zero
cosmological constant, mass or charge parameters could add more delicacy
to the theory. However, once more we note that one should avoid choosing
spacetimes with horizons or singularities, especially for the interior
of the particle. A flat Minkowski spacetime not only simplifies the
calculations, but is a great choice since it does not suffer from
such problems. Furthermore, the theory could be extended to higher-orders.
We predict that in higher orders, to have a non-trivial result, again
the Lovelock parameters of the inside and outside should be different.
As another research option, one could investigate the likelihood of
such constructions in other modified theories of gravity for which
the thin-shell formalism is already established; theories like $f\left(R\right)$
gravity. However, nothing could be more unprecedented than identifying
two rotating solutions to construct a rotating particle model. This
latter, rather than mass, charge and dimensions of the particle, could
justify the spin of the particle and somehow include quantum effects.

\section{Appendix A: Thin-shell formalism in brief (for $n=3$)}

We consider two manifolds $\mathcal{M}^{\pm}$separated by a thin-shell
at their common boundary. The line element is globally given by
\begin{equation}
ds_{\pm}^{2}=-f_{\pm}\left(r_{\pm}\right)dt_{\pm}^{2}+f_{\pm}^{-1}\left(r_{\pm}\right)dr_{\pm}^{2}+r_{\pm}^{2}\left(d\theta_{\pm}^{2}+\sin^{2}\theta_{\pm}d\phi_{\pm}^{2}\right),\label{A1}
\end{equation}
in which ($+$) refers to the outer and ($-$) to the inner regions.
The hypersurface $\Sigma:r-a\left(\tau\right)=0$, defines the shell
in which $\tau$ refers to the proper time at the shell. The induced
metric $h_{ij}^{\pm}$ for $\mathcal{M}^{\pm}$ with coordinates $\xi$
is defined by
\begin{equation}
ds_{\pm}^{2}=ds_{\Sigma}^{2}=h_{ij}^{\pm}d\xi^{i}d\xi^{j},\label{A2}
\end{equation}
where $h_{ij}^{\pm}=g_{\alpha\beta}\frac{\partial x_{\pm}^{\alpha}}{\partial\xi^{i}}\frac{\partial x_{\pm}^{\beta}}{\partial\xi^{j}}$.
Note that the indices are $\left\{ \alpha,\beta,\cdots\right\} =\left\{ t,r,\theta,\phi\right\} $
for the bulks, and $\left\{ i,j,\cdots\right\} =\left\{ \tau,\theta,\phi\right\} $
for the thin-shell. In terms of the proper time we have
\begin{equation}
\left.ds_{\pm}^{2}\right|_{r=a}=ds_{\Sigma}^{2}=-d\tau^{2}+a^{2}\left(\tau\right)\left(d\theta^{2}+\sin^{2}\theta d\phi^{2}\right),\label{A3}
\end{equation}
in which $\theta=\left.\theta_{\pm}\right|_{r=a}$ and $\phi=\left.\phi_{\pm}\right|_{r=a}$.
The normal unit vector to the shell is defined by
\begin{equation}
n_{\gamma}^{\pm}=\pm\frac{1}{\sqrt{\Delta^{\pm}}}\frac{\partial\Sigma^{\pm}}{\partial x^{\gamma}},\label{A4}
\end{equation}
where $\Delta^{\pm}$ is the normalizing factor for the unit vector
such that $n_{\gamma}^{\pm}n^{\gamma\pm}=1$ for the timelike shell.
Next, we define the extrinsic curvature tensor (the second fundamental
form) by
\begin{equation}
K_{ij}^{\pm}=-n_{\gamma}^{\pm}\left(\frac{\partial^{2}x_{\pm}^{\gamma}}{\partial\xi^{i}\partial\xi^{j}}+\Gamma_{\alpha\beta}^{\gamma\pm}\frac{\partial x_{\pm}^{\alpha}}{\partial\xi^{i}}\frac{\partial x_{\pm}^{\beta}}{\partial\xi^{j}}\right),\label{A5}
\end{equation}
where $\Gamma_{\alpha\beta}^{\gamma\pm}$ are the Christoffel symbols
compatible with the bulk metrics. In general relativity, the Einstein
equation on the shell is described by
\begin{equation}
-S_{j}^{i}=\left[K_{j}^{i}-K\delta_{j}^{i}\right]_{-}^{+},\label{A6}
\end{equation}
where $K\equiv tr\left(K_{j}^{i}\right)$, and the square brackets
denote a jump in the embraced quantity. Also, the surface stress tensor
is defined by $S_{j}^{i}=\left(\sigma,p_{\theta},p_{\phi}\right)$,
where $\sigma$ is the surface energy density and $p_{\theta}=p_{\phi}=p$
are the angular pressures. Although here we have briefly reviewed
the formalism for $n=3$, it easily could be extended to higher dimensions.
Finally, depending on which theory is adopted, the structure of $S_{j}^{i}$
may be composed of rather complicated expressions. For instance, Eqs.
(\ref{eq:27}) and (\ref{eq:42}) represent the counterparts of Eq.
(\ref{A6}) in pGB and pTOLG, respectively.

\section*{Acknowledgment}

SDF would like to thank the Department of Physics at Eastern Mediterranean
University, especially the chairman of the department, Prof. İzzet
Sakallı for the extended facilities.

\end{document}